\documentstyle[12pt]{article}
\def\al{\alpha}
\def\be{\beta}
\def\ga{\gamma}
\def\de{\delta}
\def\ep{\epsilon}

\def\th{\theta}

\def\ka{\kappa}
\def\la{\lambda}

\def\si{\sigma}

\def\Th{\Theta}
\def\La{\Lambda}

\def\mn{{\mu\nu}}

\def\frac#1#2{{\textstyle{{#1}\over {#2}}}}

\def\lsim{\mathrel{\rlap{\lower4pt\hbox{\hskip1pt$\sim$}}
    \raise1pt\hbox{$<$}}}
\def\gsim{\mathrel{\rlap{\lower4pt\hbox{\hskip1pt$\sim$}}
    \raise1pt\hbox{$>$}}}
\def\sqr#1#2{{\vcenter{\vbox{\hrule height.#2pt
         \hbox{\vrule width.#2pt height#1pt \kern#1pt
         \vrule width.#2pt}
         \hrule height.#2pt}}}}

\newcommand{\beq}{\begin{equation}}
\newcommand{\eeq}{\end{equation}}
\newcommand{\bea}{\begin{eqnarray}}
\newcommand{\eea}{\end{eqnarray}}
\newcommand{\rf}[1]{(\ref{#1})}
 
\renewenvironment{thebibliography}[1]
 { \rm
   \begin{list}{\arabic{enumi}.}
    {\usecounter{enumi} \setlength{\parsep}{0pt}
     \setlength{\itemsep}{3pt} \settowidth{\labelwidth}{#1.}
     \sloppy
    }}{\end{list}}
 
\begin{document}
\titlepage
 

\vglue 1cm
            
\begin{center}
{{\bf Yang-Mills Instantons with Lorentz Violation \\}
\vglue 1.0cm
{Don Colladay and Pat McDonald\\} 
\bigskip
{\it New College of Florida\\}
\medskip
{\it Sarasota, FL, 34243, U.S.A.\\}
 
\vglue 0.8cm
}
\vglue 0.3cm
 
\end{center}
 
{\rightskip=3pc\leftskip=3pc\noindent
An analysis is performed of instanton configurations in pure Euclidean Yang-Mills theory containing small 
Lorentz-violating perturbations that maintain gauge invariance. 
Conventional topological arguments are used to show that the general classification of
instanton solutions involving the topological charge is the same as in the standard case.
Explicit solutions are constructed for general gauge invariant corrections to the action that are quadratic in
the curvature.  The value of the action is found to be unperturbed to lowest order in the 
Lorentz-violating parameters.
}

\vskip 1 cm

\vskip 1 cm


\newpage
 
\baselineskip=20pt
 
{\bf \noindent I. INTRODUCTION}
\vglue 0.4cm

As is well known, solving pure Yang-Mills theory involves a complicated set of nonlinear 
partial differential equations.
Using a series of clever arguments, 
some exact solutions to the pure Yang-Mills field equations formulated in 
four-dimensional Euclidean space were first constructed in the mid seventies \cite{bpst}.
The complete set of finite action solutions was eventually classified using
what is now known as the ADHM construction \cite{adhm}.
Subsequently, instanton physics has stimulated much research in both
physics and mathematics \cite{review}.

In pure four-dimensional Yang-Mills theory, Lorentz symmetry and renormalizability 
coupled with gauge invariance implies that the Lagrange density 
naturally takes the form of the trace of the square of the curvature tensor.
If pure Yang-Mills theory arises as the low energy limit of some more fundamental theory,
it is possible that real physical fields obey a slightly modified version of the conventional
equations in which some of the underlying symmetries are spontaneously broken.
Specifically, Lorentz and CPT invariance, as well as gauge invariance can be
affected \cite{cpt98}.

The original motivation for this possibility arose in string theory \cite{kps},
and more recently has been analyzed within the context of
noncommutative geometry \cite{noncom}.
A framework called the Standard Model Extension (SME) incorporates general 
fundamental symmetry violations that are consistent with coordinate 
reparametrization invariance\footnote{Geometrically, the symmetry violations can
be defined using fixed sections of appropriate fibre bundles.}
within the context of quantum field theory \cite{ck}.
Usually it is convenient to restrict the full range of possible violations
to maintain certain subgroups of the original symmetry group.
For instance, translational invariance, gauge invariance 
and power-counting renormalizability are typically assumed to avoid
many of the potential inconsistencies that may arise without these assumptions.
This restriction produces a minimal version of the full SME.

The aim of this paper is to analyze the instanton solutions for a Yang-Mills action
in the presence of Lorentz violation.
The main result is that the general classification of the instanton solutions
involving the topological charge still applies when Lorentz violation is present.
In addition, the value of the Euclidean action is found to be invariant to lowest
order in the Lorentz-violating perturbations.
Specific calculations are performed within the framework of the minimal SME,
but some of the results are in fact more general.
In section II, the notation and conventions are described.
The existence of static solutions in arbitrary dimensions is 
examined in section III. 
Section IV contains the general theory of instantons with Lorentz violation,
while section V restricts to the specific example of SU(2) instantons
with unit winding number.
Section VI summarizes the results.
The appendix contains an exact solution for the perturbed instantons 
in the presence of a spatially isotropic Lorentz-violating background tensor.

\vglue 0.6cm
{\bf \noindent II. NOTATION AND CONVENTIONS}
\vglue 0.4cm

The conventions used for the Yang-Mills gauge theory are
presented in this section. 
Let G be a compact Lie group with Lie algebra L(G).
The base manifold is taken to be $M = {\bf R}^4$ and the gauge potential
components for the principle G-bundle $P \rightarrow M$ are denoted
\beq
A_\mu(x) \equiv A_\mu^a (x) L_a \quad ,
\eeq 
where the $L_a$ are hermitian generators of a Lie algebra defined by
\beq
[L_a, L_b] = i C_{abc} L_c \quad ,
\eeq
with structure constants $C_{abc}$ antisymmetric in all indices.
The normalization of the generators is fixed by imposing
\beq
Tr(L_a L_b) = \frac 1 2 \delta_{ab}
\quad .
\eeq
The associated unitary Lie group elements 
that generate gauge transformations are denoted by
\beq
U(x) = e^{-i \omega^a (x)L_a} \quad .
\eeq
These act on the gauge fields via the transformation rule
\beq
A^\mu (x) \rightarrow U(x) A^\mu (x) U^{-1}(x) - {i \over g} 
U(x) \partial^\mu U^{-1}(x)
\eeq
The field strength tensor is defined as
\beq
F^\mn = - {i \over g} [D^\mu, D^\nu] \quad ,
\eeq
where the covariant derivative is taken as 
$D^\mu = \partial^\mu + i g A^\mu$.
The field strength transforms under gauge transformations as
\beq
F^\mn \rightarrow U(x) F^\mn U^{-1}(x) \quad .
\eeq
The dual of $F$ is defined as
\beq
\tilde F_\mn \equiv \frac 1 2 \ep_{\mn\al\be} F^{\al\be} \quad ,
\eeq
where the Levi-Civita tensor is defined such that $\ep^{0123} = +1$.

In four dimensional Minkowski space, with metric $g = diag(1,-1,-1,-1)$,
The most general gauge invariant\footnote{The gauge invariance of the $k_{AF}$ 
term can be easily established for infinitesimal gauge transformations.  
Large gauge transformations may contribute nontrivially to the action.} 
and power counting renormalizable action is \cite{ck}
\bea
& & \hspace{-1cm} S_M(A) = - {1 \over 2} \int d^4 x Tr \left[ F^\mn F_\mn \right. 
+  (k_F)_{\mn \alpha \beta} F^\mn F^{\al \be}
\nonumber \\ 
& & \hspace{3cm} + \left. (k_{AF})_\ka \ep_{\ka \la \mn} 
(A^\la F^\mn - \frac 2 3 i g A^\la A^\mu A^\nu)\right]
\quad ,
\label{minkac}
\eea
where the $k_F$ and $k_{AF}$ terms are small, constant background parameters.
Gauge invariance fixes these parameters to be singlets under the action of the gauge group.
The $k_{AF}$ terms present theoretical difficulties associated with negative contributions
to the energy \cite{cfj} even in the Abelian case, 
and are therefore not considered further in this work.
On the other hand, the $k_F$ terms do not cause similar problems 
provided a concordant frame \cite{kl}, in which the parameters are small enough, is used.
The parameters $k_F$ satisfy the symmetries of the Riemann curvature 
tensor\footnote{The totally antisymmetric component would contribute
a term to the action that is proportional to the topological charge.  
This corresponds to the Lorentz-covariant
$\theta$ term used in QCD.}
with vanishing total trace.
This means that there are 19 independent coefficients that parameterize the violation.

\vglue 0.6cm
{\bf \noindent III. STATIC SOLUTIONS}
\vglue 0.4cm

In the conventional case, finite-action static solutions are ruled out in all but
four spatial dimensions by considering various integrals of the field
strength products motivated by the form of the energy momentum tensor\cite{deser}.
This result also holds in the Lorentz-violating case due to an analogous
argument that will now be presented.

The partially symmetrized energy momentum tensor arising from 
the action in Eq.~\rf{minkac}
generalized to $n$ spatial dimensions is given by the expression
\beq
\Th^\mn = 2 Tr [- F^{\nu}_{~\ga} (F^{\mu\ga} + k_F^{\mu\ga\al\be} F_{\al\be})
+ \frac 1 4 \eta^\mn F_{\al\be}(F^{\al\be}+ k_F^{\al\be\la\ka}F_{\la\ka})]
\quad ,
\eeq
and explicitly satisfies $\partial_\mu \Th^\mn = 0$. 
Choosing the static gauge in which $A^k$ is independent of time,
the following constraints on finite energy solutions can be derived
using the field equations
\beq
\int dx^n Tr F_{0k}(F^{0k} + k_F^{0k\al\be} F_{\al \be}) = 0
\quad ,
\eeq
and
\beq
(n - 4) \int dx^n Tr F_{ij} (F^{ij} + k_F^{ij \al\be} F_{\al\be}) = 0
\quad .
\label{no}
\eeq
Methods analogous to the ones presented in \cite{deser} have been applied
to obtain the above results. 
These relations imply that no static solutions 
with nonvanishing action\footnote{Nontrivial static solutions
with vanishing action are not expected provided $k_F$ is small.  
This is because $F=0$ is an extremum of the action, 
meaning that any field of order $k_F$ must be away from the extremum.}
exist when $n \ne 4$.
This result is the same as the conventional situation.

\vglue 0.6cm
{\bf \noindent IV. INSTANTON SOLUTIONS}
\vglue 0.4cm

To study the instanton solutions, the action is analytically continued to Euclidean
space using imaginary time, and a new Euclidean action $S_E \equiv -i S_M$ is defined.
The conventions used in this paper are obtained using the replacements $x^0 \rightarrow -i x_E^0$, 
$x^k \rightarrow x_E^k$, while the gauge field components
are altered to $A^0 \rightarrow i A^0_E$, and $A^k \rightarrow A^k_E$.
Each time component of $k_F$ also gets multiplied by a factor of $i$ to define
its Euclidean counterpart.
The Euclidean action becomes (dropping all $E$ subscripts)
\beq
S(A) = + {1 \over 2} \int d^4 x Tr[(F^\mn F^\mn) + (k_F)^{\mn \alpha \beta} F^\mn F^{\al \be} ]
\quad ,
\eeq 
with metric $\delta^\mn$.
The Euler-Lagrange equations of motion (for this Euclidean action) are
\beq
[D^\mu, F^{\mn} + k_F^{\mn \al \be} F^{\al\be}] = 0 \quad ,
\label{eom1}
\eeq
while the Bianchi Identity 
\beq
[D^\mu, \tilde F^{\mn}] = 0 \quad ,
\label{eom2}
\eeq
follows from the definition of $F$ in terms of the gauge potential.

The topological charge $q$ is defined as in the usual case
\beq
q = {g^2 \over 16 \pi^2}\int d^4 x Tr \tilde F^\mn F^\mn \quad ,
\eeq
and conventional arguments indicate that $q$ remains an integer,
even in the presence of Lorentz violation. 
Specifically, the identity
\beq
\frac 1 4 Tr \tilde F^\mn F^\mn = \partial^\mu X^\mu
\quad ,
\eeq
where
\beq
X^\mu \equiv \frac 1 4 \ep^{\mn\la\ka} Tr(A^\nu F^{\la\ka} - \frac 2 3 i g A^\nu A^\la A^\ka)
\quad ,
\eeq
ensures that the topological charge depends only on the pure-gauge boundary conditions
satisfied by the potential far away from the nonvanishing curvature of the instantons.
The quantity $q$ is therefore the first Pontryagin number that corresponds to the 
winding number of the map from the gauge group to the 
three-sphere at large $|x|$.  
The specific form of the action does not matter, provided that it is in fact gauge
invariant, and that finiteness of the action restricts the curvature from contributing
to the topological charge at the boundary.
This means that the properties of the topological charge should be preserved, 
even in the more general case of
the SME that includes nonrenormalizable, but gauge invariant corrections
to the pure Yang-Mills sector.  In particular, since any physical theory of noncommutative 
gauge fields is argued to be equivalent to a standard gauge theory 
in the context of the SME \cite{noncom},
the topological charge should remain integral in realistic noncommutative Yang-Mills theories.

For calculational purposes, it is convenient to introduce the quantity\footnote{
Note that $F^\prime$ is
not in general the curvature of a connection, so the theory is not automatically isomorphic
to the conventional one with $k_F = 0$.} 
\beq
F^{\prime \mn} = F^\mn + \frac 1 2 k_F^{\mn\al\be}F^{\al\be} \quad .
\eeq
The action then takes the conventional form in terms of $F^\prime$ to lowest order in $k_F$.

Consider the inequality
\beq
\frac 1 2 \int d^4 x Tr(F^\prime \mp \tilde F^\prime)^2 \ge 0 \quad .
\eeq
This implies that
\beq
S \ge \pm \frac 1 2 \int d^4 x Tr[F^\mn \tilde F^\mn + \frac 1 2 
(k_F^{\mn \al \be} + \tilde k_F^{\mn \al \be}) F^\mn \tilde F^{\al\be}]
\quad ,
\label{lowbound}
\eeq
where $\tilde k_F^{\mn \al\be} \equiv \frac 1 4 \ep^{\mn \la \ka} k_F^{\la\ka\rho\si}
\ep^{\rho \si \al \be}$ is defined as the dual to $k_F$.
The upper sign is chosen for $q > 0$ and the lower sign for
$q<0$.

The first term is proportional to the topological charge while the second term 
generates a correction to the lower bound on $S$.
Provided $k_F$ is small, the correction term will be much smaller than the 
topological charge term and the perturbed instantons will be 
close to the conventional ones.  
This implies that the general classification of the instanton
solutions in terms of the winding number will remain unaltered.

It is evident from the form of the correction to the lower bound that
splitting the coefficients $k_F$ according to their
duality properties will be useful.
This decomposition is analogous to the separation of the Riemann tensor of general relativity
into a Ricci tensor and a trace-free Weyl conformal tensor.
The anti-self-dual $k_F$ components correspond to the Ricci tensor components
while the self-dual $k_F$ terms correspond to the Weyl conformal tensor.
For the case $k_F = -\tilde k_F$,
the lower bound on the action is independent of 
continuous perturbations of $F$ that do not change the topological charge by an integer, 
and the minimum is attained for the modified duality condition 
\beq
F^\prime \simeq \pm \tilde F^\prime
\label{moddual}
\quad ,
\eeq
where the symbol $\simeq$ is used to denote an equality to lowest order in $k_F$.
To construct the perturbed solutions, the potential can be expanded
about 
the conventional ($k_F = 0$) self-dual and anti-self-dual potentials, 
denoted by $A_{SD}$ and $A_{ASD}$. 
The corresponding 
field tensors are written similarly as 
$F_{SD}$, and $F_{ASD}$.
It remains to show that solutions to the modified duality condition that are consistent with
the Bianchi identity exist.
The anti-self duality condition on $k_F$ implies that it must be of the form
\beq
k_F^{\mn \al \be} = \La_{k}^{[\mu[\al} \de^{\nu] \be]}
\quad ,
\eeq
where $\La_k^{\mn} = \frac 1 2 k_F^{\al\mu\al\nu}$ is a traceless, symmetric matrix
that depends on the trace-components of $k_F$.
In fact, the explicit solution can be guessed since the form of the correction 
to the action is related to the conventional action as described in 
the skewed coordinate system $\tilde x^\mu \equiv x^\mu + \La^{\mn}_k x^\nu$.
These terms are exactly the ones that may be transferred to other sectors
using an appropriate field redefinition \cite{cm}, so it is not surprising that 
they yield a conventional version of pure Yang-Mills theory when described in 
skewed coordinates.
Note that this does not imply the absence of physical effects arising from an anti-self-dual
$k_F$ term in the action. 
Redefining coordinates effects all fields, not just the Yang-Mills gauge
potential, so if the instantons are expressed in terms of new coordinates,
the Lorentz-violation will show up in the Lagrangian for other particle species
that are coupled to the instantons.

The explicit form for the perturbed self-dual instanton gauge potentials are given by
\beq
A_+^\mu(x) \simeq A^\mu_{SD}(\tilde x) + \La_k^{\mn} A^\nu_{SD}(x)
\quad ,
\label{asdinst}
\eeq
yielding a perturbed field tensor 
\beq
F_+^{\mn} \simeq F_{SD}^\mn (\tilde x) 
- \La_k^{[ \mu \al} \de^{\nu ] \be } F^{\al\be}_{SD} (x)
\simeq F_{SD}^\mn (\tilde x) - \frac 1 2 k_f^{\mn\al\be} F^{\al\be}_{SD} (x)
\quad ,
\eeq
that satisfies the modified duality condition \rf{moddual}.
Note that the approximation is in fact not necessary in this case, 
but the notation becomes rather cumbersome for general $k_F$.
The exact solution for the O(3) rotationally invariant component of $k_F$
(which is in fact anti-self-dual) is presented in the appendix.

Next, the case $k_F = \tilde k_F$ is considered.
This condition implies that $k_F$ has the symmetries of the Weyl conformal
tensor with vanishing single traces.
In this case, the simple argument given above for anti-self-dual $k_F$ fails 
because the lower bound on the action in Eq.~\rf{lowbound}
is not a topological invariant, and is therefore sensitive to small perturbations 
in the field strengths.
In this case, there is no obvious duality condition and the equations of
motion must be solved directly for the perturbed instanton solutions.
A solution to lowest order in $k_F$ always exists, since the equations reduce
to a set of linear, second order elliptic partial differential equations for the gauge fields.
The propagators for spin-1 particles in instanton background fields have been
previously constructed \cite{bccl} and are exactly what is needed to formally solve the equations.
An explicit example is presented in the next section.

For general $k_F$, the perturbed field strength may be written as a small perturbation of 
either the $F_{SD}$ or the $F_{ASD}$ solutions.
Remarkably, the approximate value of the action is the same as the conventional
case.  For example, an instanton that is close to self-dual yields an action of
\beq
S \simeq \frac 1 2 \int d^4 x Tr (F^2 + k_F^{\mn\al\be}F^\mn_{SD} F^{\al\be}_{SD})
\quad .
\eeq
The first term is the conventional action and is invariant to lowest order in any
perturbation of the fields due to the fact that the action is at an extremum for the
self dual solutions.
The O(4) symmetry of the conventional self-dual solutions imply that the second term must
vanish, since only observer Lorentz-invariant components of $k_F$ can contribute after
the trace is performed.
These terms are zero due to the Lorentz-violating nature of $k_F$.
The same arguments apply to the instantons that are close to the anti-self-dual solutions.
The numerical value of the action to leading order in $k_F$ is therefore given by
the conventional formula
\beq
S \simeq (8 \pi^2/g^2) |q| 
\quad ,
\eeq
for the general case involving arbitrary $k_F$ values.
This argument can also be generalized to 
nonrenormalizable corrections to the pure Yang-Mills sector involving
powers of the curvature tensor.
This works because any higher order Lorentz-violating corrections
must vanish when the O(4) symmetric solutions are substituted into
the action.
As mentioned previously, any realistic theory of noncommutative gauge fields 
is argued to be equivalent to 
a subset of the SME \cite{noncom}, therefore it can be inferred that
any realistic theory of noncommutative Yang-Mills fields
should not affect the value of the Euclidean action for the instantons 
to lowest order in the noncommutative, Lorentz-violating $\theta^\mn$ parameters. 

\vglue 0.6cm
{\bf\noindent V. Instantons in SU(2)}
\vglue 0.4cm

To analyze instanton structure, an explicit map is constructed from
the asymptotic three sphere $S^3$ of Euclidean space
into the Yang-Mills
gauge group $G$.
The winding number of this map determines the topological charge and therefore 
the general instanton structure according to the lower bound of the action in
Eq~\rf{lowbound}.
For any simple Lie group $G$, a theorem by Bott \cite{Bott}
proves that any mapping of $S^3$ into $G$ can be continuously deformed into
a mapping into an
SU(2) subgroup of $G$.
It is therefore sufficient to fix SU(2) as the gauge group to 
construct explicit solutions that will exhibit the generic effect
of Lorentz violation on the instanton structure.

Here we work with the explicit solutions for $q=1$, or unit topological charge.
The conventional solutions are denoted using the self-dual, antisymmetric tensor
$\tau^\mn$, where $\tau^{0i} \equiv \si^i$ and $\tau^{ij} \equiv \ep^{ijk} \si^k$,
in terms of the conventional Pauli matrices $\si^i$.
This definition provides an explicit embedding of 
SU(2)$\rightarrow$SU(2)$\times$SU(2) which is isomorphic to O(4).
The commutation relations 
\beq
[\tau^\mn, \tau^{\al\be}] = 2i(\de^{\mu\al} \tau^{\nu\be} - \de^{\mu\be}\tau^{\nu\al}
- \de^{\nu\al}\tau^{\mu\be} + \de^{\nu\be} \tau^{\mu\al} ) \quad ,
\eeq
and trace relations
\beq
Tr(\tau^\mn\tau^{\al\be}) = 2(\de^{\mu\al}\de^{\nu\be} - \de^{\mu\be} \de^{\nu\al}
+ \ep^{\mn\al\be})
\quad ,
\eeq
follow from the above definition.
These quantities may also be expressed using the relation
$\tau^\mn = i (\tau^\nu {\tau^\mu}^\dagger - \delta^\mn)$,
where $\tau^\mu \equiv (i, \vec \si)$.
The self-dual gauge field corresponding to $q=1$ can be expressed as
\beq
A^\mu_{SD} = - {\tau^{\mn}x^\nu \over g (\rho^2 + x^2)}
\quad ,
\eeq
and the associated field strength is
\beq
F^{\mn}_{SD} = {2 \rho^2 \over g (\rho^2 + x^2)^2} \tau^\mn
\quad .
\eeq
The parameter $\rho$ determines the instanton size,
while the center of the instanton is taken to be at the origin for simplicity.
The anti-self-dual solutions are the parity transform of the above fields.
These can be expressed using $\tilde x = (x^0,-\vec x)$ as
$A^0_{ASD}(x) = A^0_{SD}(\tilde x)$, 
$A^i_{ASD}(x) = - A^i_{SD}(\tilde x)$, 
$F^{0i}_{ASD}(x) = - F^{0i}_{ASD}(\tilde x)$, and
$F^{ij}_{ASD}(x) = F^{ij}_{ASD}(\tilde x)$.
This transformation may also be implemented by the transformation
$\tau^\mn \rightarrow \overline \tau^\mn$ defined by
$\tau^{0i} \rightarrow \overline \tau^{0i} = - \tau^{0i}$, 
and $\tau^{ij} \rightarrow \overline \tau^{ij} = \tau^{ij}$.
A useful expression for this quantity is 
$\overline \tau^{\mn} = i({\tau^\nu}^\dagger \tau^\mu - \delta^\mn)$.

For the case $k_F = - \tilde k_F$, the modified solutions have already
been expressed using Eq.\rf{asdinst} and do not require more explicit
computation.
For the case $k_F = \tilde k_F$, the field equations \rf{eom1} and 
\rf{eom2} must be solved
directly since no obvious duality condition can be determined from
Eq~\rf{lowbound} due to the non-invariant lower bound.
To accomplish this, the vector potential is expanded as a 
perturbation of the self-dual\footnote{Only the solution that
is close to self-dual is presented here for notational simplicity,
the close to anti-self-dual solution may be constructed using an analogous 
procedure.} 
solution $A = A_{SD} + A_k$
and the linear terms in $A_k$ are retained in the equation of motion.
The Bianchi identity \rf{eom2} is automatically satisfied 
and the equations of motion \rf{eom1} become 
(in the Lorentz gauge $\partial^\mu A^\mu = 0$)
\beq
[D_{SD}^\nu,[D_{SD}^\nu, A_k^\mu]] + 2 i g [F_{SD}^\mn, A_k^\nu] 
- i g [D_{SD}^\mu, [A^\nu_{SD}, A^\nu_k]]= j_k^\mu
\quad ,
\label{eom}
\eeq
where
\beq
j_k^\mu \equiv k_F^{\mn\al\be} [D^\nu_{SD}, F_{SD}^{\al\be}]
\quad ,
\eeq
and $D^\mu_{SD} \equiv \partial^\mu + i g A^\mu_{SD}$ is the covariant 
derivative in the conventional self-dual instanton background.

This equation can be solved by performing a convolution of $j_k$ with the corresponding
propagator for spin-1 particles in an external instanton field.
This propagator has been formally constructed \cite{bccl}, but the explicit
form is rather unwieldy and cannot be easily expressed analytically.
An alternative approach is adopted here that uses a combination of the 
propagator approach and a direct substitution technique.
First, the solution is studied to lowest
order in $\rho^2/x^2$, corresponding to the asymptotic region far from the self-dual 
instanton curvature density.
This provides the general tensorial structure of the instanton correction
that serves as an ansatz for general values of $x^2$, generating
a simple form for the solution to the problem.

It is convenient to perform a gauge transformation to the singular gauge
using $U(x) = - i x \cdot \tau^\dagger / x$ so that the potential is better
behaved for large $x$.
The transformed potential becomes 
\beq
\overline A_{SD}^\mu = - {\rho^2 \overline \tau^{\mn} x^\nu \over g x^2 (\rho^2 + x^2)}
\quad ,
\eeq
with associated field strength
\beq
\overline F_{SD}^{\mn} = {4 \rho^2 \over g (\rho^2 + x^2)^2}\overline \tau^{[\mu\al}
(\frac 1 4 \delta^{\nu]\al} - {x^{\nu]} x^\al \over x^2})
\quad .
\eeq
In this gauge, the transformed $\overline j_k$ is
\beq
\overline j_k^\mu = {48 \rho^2 \over g x^2 (\rho^2 + x^2)^3}
k_F^{\mn\al\be} \overline \tau^{\al \ga}I^{\ga \nu \be}(x)
\quad ,
\eeq
where
\beq
I^{\ga\nu\be} \equiv x^\ga x^\nu x^\be - \frac 1 6 x^2 
(\de^{\nu\ga} x^\be + \de^{\ga\be} x^\nu + \de^{\be \nu} x^\ga)
\quad ,
\eeq
is a totally symmetric tensor.

The advantage of working in the singular gauge is that
the above expressions are all quadratic in $\rho$.
This means that to lowest order in $\rho^2 / x^2$, the propagator
may be approximated by the free field Green's function
\beq
G_0(x,y) = {1 \over 4 \pi^2 (x - y)^2}
\quad ,
\eeq
satisfying $\partial^\mu \partial^\mu G_0 = - \delta^{(4)} (x-y)$.
The perturbed potential to lowest order in $\rho^2/x^2$ (in the singular gauge) 
is then given by 
\beq
\overline A^\mu_k \simeq - \int d^4 y G_0(x,y) \overline j_k^\mu(y)
\quad .
\eeq
This integral can be performed using standard field theoretic integration
techniques.
The result of the computation is
\beq
\overline A^\mu_k \simeq - {4 \rho^2 \over g x^6} 
k_F^{\mn \al \be} \overline \tau^{\al\ga} I^{\ga\nu\be}(x)
\quad .
\label{smallx}
\eeq
It can be seen that the tensorial structure of $\overline j_k$
has been preserved by the convolution with $G_0$.
Some complications arise due to divergent logarithms that cancel out in the computation,
but these do not cause theoretical difficulties because
the validity of the solution can be verified by direct substitution into the equation
of motion.
It remains to check that the Lorentz gauge condition is satisfied
by this solution.
Direct calculation shows that this is the case provided $k_F$ is self-dual, 
the current case of interest.
This indicates that
this solution method works for the terms that cannot be removed using
a reparametrization of the coordinates.

For general values of $x^2$, an unknown scalar function is 
included in the expression \rf{smallx} to produce an ansatz of the form
\beq
\overline A^\mu_k = - {4 \rho^2 \over g} f(x^2) 
k_F^{\mn \al \be} \overline \tau^{\al\ga} I^{\ga\nu\be}(x)
\quad .
\eeq 
Remarkably, upon substitution into the equation of motion \rf{eom},
the tensorial structure factors out and
the following second order linear differential equation is found for $f$
\beq
x^4 (\rho^2 + x^2) f^{\prime\prime} + 5 x^2 (\rho^2 + x^2) f^\prime
+ 3 \rho^2 f = - {3 \over (\rho^2 + x^2)^2 }
\quad .
\label{equf}
\eeq
This equation has a regular singular point at $x = 0$,
causing the homogenoeous solutions to both be badly behaved at the
origin. 
Moreover, any contribution to the homogeneous equation of motion
would correspond to a solution to the conventional equations of 
motion in an instanton background and is therefore not of interest 
in the present context.
On the other hand, the particular solution is well-behaved at the
origin as can be verified using the following series expansion for 
$f$ about $x = 0$
\beq
f(x^2) = {1 \over \rho^6} \sum^\infty_{n = 0} 
a_n \left({x^2 \over \rho^2}\right)^n
\quad ,
\eeq
and expanding the right hand side of Eq.~\rf{equf} as
\beq
- {3 \over (\rho^2 + x^2)^2 } = - {3 \over \rho^4} 
\sum^\infty_{n = 0} (-1)^n (n + 1) \left(  {x^2 \over \rho^2}\right)^n
\quad ,
\eeq
valid for $x^2 / \rho^2 < 1$.
The resulting recursion relation for the $a_n$ coefficients is
\beq
a_{n + 1} = {3 (-1)^n \over n + 4} - {n \over n + 2} a_n
\quad ,
\eeq
with $a_0 = -1$.
The first few terms gives 
\beq
f(x^2) \approx 1/\rho^6(-1 + {3 \over 4} \left(  {x^2 \over \rho^2}\right) 
- {17 \over 20}\left(  {x^2 \over \rho^2}\right)^2 
+ {37 \over 40} \left(  {x^2 \over \rho^2}\right)^3 - \cdots )
\quad ,
\eeq
demonstrating the finite behavior near the origin.
For large $x^2$,
a similar expansion in $\rho^2 / x^2$ shows that the function
approaches $f(x^2) \rightarrow 1/x^6$ as expected.

Transforming the perturbed potential back to the regular gauge yields
\beq
A_k^\mu \simeq {2 \rho^2 x^2 \over 3 g} f(x^2) k_F^{\mn\al\be} \tau^{\al\ga} 
(\de^{\ga \nu} x^\be + \de^{\be \ga} x^\nu - \de^{\nu\be}x^\ga)]
\quad ,
\eeq
verifying that $A_k$ is zero at the origin in the regular gauge as is required
by continuity of the gauge field.
The perturbation term behaves asymptotically as $\sim 1/x^3$, and therefore explicitly
does not contribute to the the topological charge as expected.
The resulting correction to the curvature can be computed,
however the specific form is not particularly illuminating.
Specifically, there seems to be no obvious generalized duality condition
satisfied by $F$ analogous to the situation for $k_F = - \tilde k_F$. 

\vglue 0.6 cm
{\bf \noindent VI. SUMMARY}
\vglue 0.4 cm

Instantons have long been studied for systems obeying strict Lorentz invariance.
In this paper, the structure of Yang-Mills instantons in the presence of small 
Lorentz-violating
background fields that maintain gauge invariance is studied for the first time.
No new nonzero action static solutions are present in $n \ne 4$ spatial dimensions
as is apparent from Eq~\rf{no}. 
The gauge invariance ensures that the conventional pure-gauge asymptotic behavior
maintains the same general structure as in the conventional case.  
This means that conventional arguments can be applied to deduce the quantization
of the topological charge.
The generality of the SME can then be exploited to infer similar results regarding
realistic noncommutative gauge theories.

Specific perturbed instanton solutions for the action considered in this paper 
are split into two categories
that depend on the duality properties of the Lorentz-violating background tensor.
For the anti-self-dual $k_F$ case, a reparametrization of the coordinates can be used to
construct deformed instantons that satisfy a modified duality condition.
The perturbed theory is isomorphic to the conventional Yang-Mills theory in this case
so the instanton structure is also isomorphic.
The O(3) rotationally invariant term of this class is worked out exactly in the appendix.

When $k_F$ is self-dual, the conventional lower bound argument involving the action
fails and the equations of motion must be solved directly.
To lowest order in $k_F$, the resulting equations are linear in the correction to
the vector potential and can be formally solved using the Euclidean propagator for a spin-1
particle in an instanton background.
For explicit calculation, it turns out to be more practical to first deduce the general tensorial
structure in the asymptotic region, then generalize the solution to arbitrary position.  
General arguments imply that the action is unaltered to lowest order in $k_F$,
but it can be seen from the exact solution given in the appendix 
that higher order corrections are in general nonzero.

\appendix
\vglue 0.6 cm
{\bf \noindent APPENDIX: EXACT SOLUTION FOR O(3) SYMMETRIC CASE}
\vglue 0.4 cm

\renewcommand{\theequation}{A\arabic{equation}}
\setcounter{equation}{0}

In this appendix, an exact solution (all orders in $k_F$) for the case of spatial
rotationally invariant $k_F$ is presented.
In this case, the tensor $k_F$ can be expressed 
in terms of one independent parameter $\tilde \ka$ as
\beq
k_F^{0i0j} = - k_F^{i00j} = - k_F^{0ij0} = k_F^{i0j0} =
- {\tilde \ka \over 2} \delta^{ij}
\quad , 
\eeq
and
\beq
k_F^{ijkl} = \tilde \ka (\de^{ik} \de^{jl} - \de^{il} \de^{jk})
\quad .
\eeq
It is convenient to introduce the notation $\tilde \ka = \sin{2 \th}$ and the
action takes the form
\beq
S = \frac 1 2 \int d^4 x Tr[F^{\mn} F^{\mn} + \sin{2 \theta} (F^{ij} F^{ij} - 2 F^{0i} F^{0i})]
\quad .
\eeq
To construct the analog of the conventional self-dual solution,
consider the following inequality
\beq
\frac 1 2 \int d^4 x Tr \{ 2[\cos{\th}F^{0i}_- - \sin{\th}F^{0i}_+]^2
+ [\cos{\th}F^{ij}_- + \sin{\th}F^{ij}_+]^2 \} \ge 0
\quad ,
\eeq
with $F^{\mn}_\pm \equiv F^{\mn} \pm \tilde F^{\mn}$.
This can be rearranged to give the relation
\beq
S \ge {8 \pi^2 \over g^2} q \cos{2 \th}
\quad . 
\eeq
The inequality is saturated when
\beq
\tilde F^{0i} = {1 - \tan{\th} \over 1 + \tan{\th}} F^{0i}
\quad ,
\eeq
and
\beq
\tilde F^{ij} = {1 + \tan{\th} \over 1 - \tan{\th}} F^{ij}
\quad .
\eeq
A solution to these equations with $q=1$ is provided by the gauge potential
\beq
A^0 = (1 + \tan{\th}) A^0_{SD}(\tilde x)
\quad , \quad 
A^i = (1 - \tan{\th}) A^i_{SD} (\tilde x)
\eeq
where $\tilde x^\mu \equiv ((1 + \tan{\th}) x^0 , (1 - \tan{\th}) x^i)$.
The resulting field strength is
\beq
F^{0i} = (1 - \tan^2{\th}) F^{0i}_{SD}(\tilde x)
\quad ,\quad 
F^{ij} = (1 - \tan{\th})^2 F^{ij}_{SD}(\tilde x)
\quad .
\eeq
The value of the resulting action can be computed directly from the curvature, 
yielding the expected value 
\beq
S = {8 \pi^2 \over g^2} q \cos{2 \th}
\quad .
\eeq
In fact, this construction applies to any conventional instanton solution,
since the spatially rotational invariant $k_F$ term corresponds to a shift in the
speed of light for the gauge fields.
It is therefore possible to construct the above solutions by rescaling the
time and spatial coordinates appropriately.
Note that this does not mean that observable effects are absent, since
interactions between the instantons and other particles with conventional
Lorentz properties may lead to physical effects.
The action is reduced relative to the conventional case by a factor of 
$\cos{\theta}$ which is 
in fact second order in the $k_F$ coefficients.
This result is in agreement with general arguments stating that the 
numerical value of the action is stable to a lowest order perturbation in $k_F$.

\vglue 0.6cm
{\bf\noindent REFERENCES}
\vglue 0.4cm


\begin{thebibliography}{xx}

\bibitem{bpst}
A. Belavin, A. Polyakov, A. Schwartz, and Y. Tyupkin,
Phys. Lett. {\bf 59B} 85 (1975).

\bibitem{adhm}
M.F. Atiyah, N.J. Hitchin, V. G. Drinfeld, and Yu. I. Manin,
Phys. Lett. {\bf 65A} 285 (1978).

\bibitem{review}
For example, see {\it Instantons in Gauge Theories}, ed. M. Shifman,
World Scientific, Singapore (1994); D. Freed and K. Uhlenbeck,
{\it Instantons and four-manifolds},
New York, Springer-Verlag (1991).

\bibitem{cpt98}
For a summary of recent theoretical models and
experimental tests
see, for example,
{\it CPT and Lorentz Symmetry}, V.A.\ Kosteleck\'y, ed., 
World Scientific, Singapore, 1999; 
{\it CPT and Lorentz Symmetry II}, V.A.\ Kosteleck\'y, ed.,
World Scientific, Singapore, 2002.

\bibitem{kps}
V.A.\ Kosteleck\'y and S.\ Samuel,
Phys.\ Rev.\ D {\bf 39}, 683 (1989);
{\it ibid.} 
{\bf 40}, 1886 (1989);
Phys.\ Rev.\ Lett.\ {\bf 63}, 224 (1989);
{\it ibid.} 
{\bf 66}, 1811 (1991);
V.A.\ Kosteleck\'y and R.\ Potting,
Nucl.\ Phys.\ B {\bf 359}, 545 (1991);
Phys.\ Lett.\ B {\bf 381}, 89 (1996);
Phys.\ Rev.\ D {\bf 63}, 046007 (2001); 
V.A.\ Kosteleck\'y, M.\ Perry, and R.\ Potting,
Phys.\ Rev.\ Lett.\ {\bf 84}, 4541 (2000). 

\bibitem{noncom}
S. Carroll, {\it et al.}, Phys.\ Rev.\ Lett.\ {\bf 87},
141601 (2001).

\bibitem{ck} 
D.\ Colladay and V.A.\ Kosteleck\'y,
Phys.\ Rev.\ D {\bf 55}, 6760 (1997);
Phys.\ Rev.\ D {\bf 58}, 116002 (1998).

\bibitem{cfj}
S. M. Carroll, G. B. Field, and R. Jackiw,
Phys.\ Rev.\ D {\bf 41}, 1231 (1990).

\bibitem{kl}
For a more detailed discussion of concordant frames, see
V. A. Kostelecky and R. Lehnert,
Phys.\ Rev.\ D {\bf 63} 065008 (2001).

\bibitem{deser}
S. Deser,
Phys.\ Lett.\ {\bf 64}, B463 (1976).

\bibitem{cm}
D. Colladay and P. McDonald,
J. Math. Phys. {\bf 43} 3554 (2002);
A. Kosteleck\'y and M. Mewes,
Phys.\ Rev.\ D {\bf 66}, 056005 (2002).

\bibitem{Bott}
R. Bott,
{\it Bull. Sos. Math. France}
{\bf 84}, 251 (1956). 

\bibitem{bccl}
L. S. Brown, R. Carlitz, D. Creamer, and C. Lee,
Phys.\ Rev.\ D {\bf 17}, 1583 (1978);
H. Levine and L. Yaffe,
Phys.\ Rev.\ D {\bf 19}, 1225 (1979).

\end{thebibliography}
\end{document}